\documentclass[aps,prc,twocolumn,superscriptaddress]{revtex4-1}  
\usepackage{amssymb,amsmath,mathtools,color}
\usepackage{graphicx}
\usepackage{color}

\usepackage{graphicx}

\begin{document}

\title{Comments on the ${}^{12}{\rm C}$-${}^{12}{\rm C}$ fusion S*-factor}         
\author{A. M. Mukhamedzhanov}
\email[Correspondence author: ]{akram@comp.tamu.edu}
\affiliation{Cyclotron Institute, Texas A$\&$M University, College Station, TX 77843, USA}
\author{X. Tang}
\email[Correspondence author: ]{xtang@impcas.ac.cn}
\affiliation{Institute of Modern Physics, Chinese Academy of Sciences, Lanzhou 730000, China}
\affiliation{Joint department for nuclear physics, Lanzhou University and Institute of Modern Physics, Chinese Academy of Sciences, Lanzhou 730000, China}
\author{D. Y. Pang}
\email[Correspondence author: ]{dypang@buaa.edu.cn}
\affiliation{School of Physics and Nuclear Energy Engineering, Beihang University, Beijing, 100191, China}
\affiliation{Beijing Key Laboratory of Advanced Nuclear Materials and Physics, Beihang University, Beihang University, Beijing, 100191, China}
\date{\today}          

\begin{abstract}
The goal of this Comment is to draw attention of the readers that the results in the published letter [A Tumino {et al,}, Nature {\bf 557}, 687 (2018)] regarding ${}^{12}{\rm C}-{}^{12}{\rm C}$ fusion  are not correct.
\end{abstract}

\maketitle

The first application of the indirect Trojan horse method (THM) was published recently in Nature to determine the astrophysical S*-factor and reaction rate of the  ${}^{12}{\rm C}$-${}^{12}{\rm C}$ fusion, which is one of the important astrophysical reactions. The THM is a powerful indirect method, which allows one to measure the S*-factors at stellar energies due to the absence of the Coulomb barrier in the ${}^{12}{\rm C}$-${}^{12}{\rm C}$ system in the THM reaction. The obtained S*-factors in \cite{NatureTHM1} demonstrate a profound rise at the ${}^{12}{\rm C}$-${}^{12}{\rm C}$  relative energy below $2.7$ MeV,  exceeding all the existing low-energy extrapolations based on the direct data by several orders of magnitudes. As a result, the reaction rate was increased by a factor of nearly 1000 around T$_9$=0.2.

However, we found at least three major problems in this work which led us to question the conclusion above. 

1) The plane-wave approach used in \cite{NatureTHM1} was developed by one of us (A.M.M.) and is valid only for the THM reactions above the Coulomb barrier and smaller charges.  However, at the 30 MeV incident energy of ${}^{14}$N, the d-${}^{24}$Mg Coulomb interaction for the ${}^{12}$C transfer reaction plays a crucial role. Completely neglecting the Coulomb interaction in \cite{NatureTHM1} led to  misleading results. A general theory \cite{muk2011}, which includes the Coulomb interactions in the three-body approach, should be used in the analysis of the THM reaction to determine the S*- factor of the ${}^{12}{\rm C}$-${}^{12}{\rm C}$ fusion rather than the plane-wave approach.  Results of the application of this general theory disapprove the steep rise of the S*-factors at low energies as reported in \cite{NatureTHM1} and will be published elsewhere. Moreover,  the final-state three-body Coulomb  interaction changes the location of the observed  in the THM resonances compared  to the ones measured in direct experiments. 
 
2) The overlap with the reliable data obtained from the direct measurements is essential for the success of the THM approach. The normalization to the direct data \cite{spillane, mazarakis1973, high, kettner, barron2006} was done in the range of $2.50-2.63$ MeV for the ${}^{20}$Ne +$\alpha_{1}$ channel. Within the direct data sets used for the normalization, the measurement from \cite{mazarakis1973} was proved to be wrong in its energy calibration, and the rising trend in the S*-factor disappears after the correction \cite{kettner, barnes}. Another direct measurement \cite{barron2006} used a smoothing method to wash out all the resonances. Without a confirmation of the correct energies of the resonances, this data set should not be included.  The $\gamma$-ray measurements reported in \cite{spillane, high, kettner} are the cross sections for the $\gamma$-transition from the first excited state to the ground state. It is inappropriate to compare the ${}^{12}$C(${}^{12}$C,$\alpha_1$)${}^{20}$Ne data with these direct data without correcting the decay branching ratio. 

3) The resonances used in the R-matrix analysis in \cite{NatureTHM1} are questionable. The feature of identical bosons in the ${}^{12}$C-${}^{12}$C entrance channel restricts the resonance spins only to be even numbers with positive parity. An arbitrary inclusion of all the ${}^{24}$Mg resonances with different parities in the R-matrix analysis violates the fundamental principle of the quantum mechanics. Within the overlapping energy range between the THM data and the direct data, there is only one peak (E$_{c.m.}$=2.56 MeV) in the ${}^{20}$Ne+$\alpha_{1}$ channel with a wrong J$^{\pi}$ of 3$^{-}$. Therefore, it is not feasible to verify whether the resonances reported in \cite{NatureTHM1} correspond to the molecular ${}^{12}$C-${}^{12}$C resonances or not. 

\section{Acknowledgments}
A. M. M. acknowledges the support by the U.S. DOE Grant No. DE-FG02-93ER40773, NNSA Grant No. DE-NA0003841 and U.S. NSF Award No. PHY-1415656 in developing the THM theory with Coulomb effects. X.D.T. acknowledges support from the National Natural Science Foundation of China under Grants No. 11021504, No. 11321064, No. 11475228, and No.
11490564, 100 talents Program of the Chinese Academy of Sciences. D.Y.P. acknowledges the support by NSFC Grant Nos. 11775013 and U1432247.

\end{document}